\documentclass[a4paper,11pt]{article}
\usepackage{pos}

\xdefinecolor{mygreen}{rgb}{0,0.35,0}
\newcommand{\blue}{\textcolor{blue} }
\newcommand{\red}{\textcolor{red} }
\newcommand{\purple}{\textcolor{purple} }
\newcommand{\orange}{\textcolor{orange} }
\newcommand{\green}{\textcolor{mygreen} }

\newcommand{\be}{\begin{equation}}
\newcommand{\bfr}{\begin{frame}}
\newcommand{\ee}{\end{equation}}
\newcommand{\efr}{\end{frame}}
\newcommand{\nn}{\nonumber}
\newcommand{\pa}{\partial}
\newcommand{\ba}{\begin{eqnarray}}
\newcommand{\ea}{\end{eqnarray}}

\usepackage{amsmath}
\usepackage{amsfonts}

\title{Finite-temperature effects on D-meson properties}

\author*[a]{Juan M. Torres-Rincon}
\author[b]{Gl\`oria Monta\~na}
\author[b]{\`Angels Ramos}
\author[c,d,e,f]{Laura Tolos}

\affiliation[a]{Institut f\"ur Theoretische Physik, Goethe Universit\"at Frankfurt, Max von Laue Strasse 1, 60438 Frankfurt, Germany}
\affiliation[b]{Departament de F\'isica Qu\`antica i Astrof\'isica and Institut de Ci\`encies del Cosmos (ICCUB), Facultat de F\'isica,  Universitat de Barcelona, Mart\'i i Franqu\`es 1, 08028 Barcelona, Spain}
\affiliation[c]{Institute of Space Sciences (ICE, CSIC), Campus UAB, Carrer de Can Magrans, 08193, Barcelona, Spain}
\affiliation[d]{Institut d'Estudis Espacials de Catalunya (IEEC), 08034 Barcelona, Spain}
\affiliation[e]{Faculty of Science and Technology, University of Stavanger, 4036 Stavanger, Norway}
\affiliation[f]{Frankfurt Institute for Advanced Studies, Ruth-Moufang-Str. 1, 60438 Frankfurt am Main, Germany}

\emailAdd{torres-rincon@itp.uni-frankfurt.de}

\abstract{We study the spectroscopy and transport properties of charmed mesons in a thermal medium by applying an effective field theory based on chiral and heavy-quark symmetries in the imaginary time formalism. Relying on unitarity constraints and self-consistency we extract the in-medium properties (masses and widths) of $D$ and $D_s$ mesons and their interactions with light hadrons. We report our findings on 1) dynamically generated states, 2) thermal evolution of chiral partners, 3) in-medium scattering amplitudes, and 4) transport coefficients below the chiral restoration temperature.}

\FullConference{%
  *** 10th International Workshop on Charm Physics (CHARM2020), ***\\
  *** 31 May - 4 June, 2021 ***\\
  *** Mexico City, Mexico - Online ***
}

\begin{document}
\maketitle

\section{Introduction}

In our contribution to the CHARM2020 workshop we have presented our latest results on the properties of $D$ mesons and other charmed states modified by the presence of a thermal medium.

These states---containing a single charm valence quark---have large masses in comparison to other scales in the system, thus behaving as Brownian particles propagating in a thermal bath (itself composed by light hadrons such as $\pi, K, {\bar K}$ and $\eta$ mesons)~\cite{Aarts:2016hap,Prino:2016cni,Dong:2019byy,Dong:2019unq}. Such scale separation helps to organize the expansion of the effective field theory, and also to simplify the kinetic equation which describes the time evolution of $D$ mesons. 

The vacuum interactions of $D/D_s$ mesons (also $D^*/D_s^*$ mesons) with the light degrees of freedom has been explored in a number of works exploiting both chiral and heavy-quark spin-flavor symmetries as guiding principles to construct the effective Lagrangian (see our works~\cite{Montana:2020lfi,Montana:2020vjg} and the references therein). By matching to recent lattice-QCD calculations, the unknown parameters of the Lagrangian can be entirely fixed, and the theory becomes predictive (but limited to the energy region for which the effective approach remains strictly valid). To describe physical hadron interactions around the energies of interest, i.e. at the resonance region, one should perform the ``unitarization'' of the scattering-matrix elements. Such a method forces exact unitarity constraints on the scattering amplitudes, and is able to generate several resonances and bound states in the complex-energy plane. Several of them are in fact compatible with states listed in the Particle Data Group review~\cite{ParticleDataGroup:2018ovx}.

In the context of relativistic heavy-ion collisions, charm quarks are produced during the initial stage of the collision (via hard scatterings) and they become part of the so-called quark-gluon plasma phase, following a fast thermalization process. After hadronization, $D$ mesons (including not only the ground states, but also all possible excitations, resonances and bound states) become part of the hadronic medium as ``heavy impurities''. For them a Brownian description turns out to be very convenient. To describe their dynamics, one needs to extend the effective approach to finite temperature, as the backreaction of the medium might become important. In addition, to fully characterize the relaxation of these Brownian particles, is it necessary to tackle their transport properties, like thermal widths, and other transport coefficients.

In this proceeding we present a summary of our results along these lines: finite temperature corrections to the dynamics of $D$ mesons and their dynamically generated states (thermal masses and decay widths), and the determination of a consistent kinetic theory together with the calculation of the relevant transport coefficients at finite temperature. Full details are given in our works~\cite{Montana:2020lfi,Montana:2020vjg,Torres-Rincon:2021yga}.

\section{Thermal effective field theory for $D$ mesons}

  The effective field theory describing $D$ mesons, and their interactions with light degrees of freedom ($\pi, K, \bar{K}, \eta$, denoted generically as $\Phi$) is described in Refs.~\cite{Montana:2020lfi,Montana:2020vjg}. In those works the approach is also extended to finite temperature using the imaginary-time formalism. It follows closely the description in vacuum and it is based on both chiral (accounting for the pseudoscalar octet of light mesons $\Phi$) and heavy-quark spin-flavor symmetry (for the states with charm quantum number). It should be noted that the free parameters (low-energy constants and cut-off regulator) are fixed already in vacuum, being the extension to finite $T$ straightforward.
  
\begin{figure}[ht]
  \centering
  \includegraphics[width=75mm]{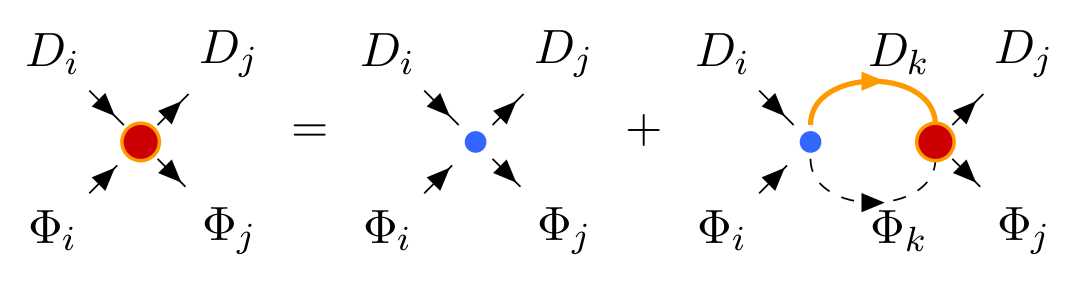}
 \hspace{8mm}
  \includegraphics[width=60mm]{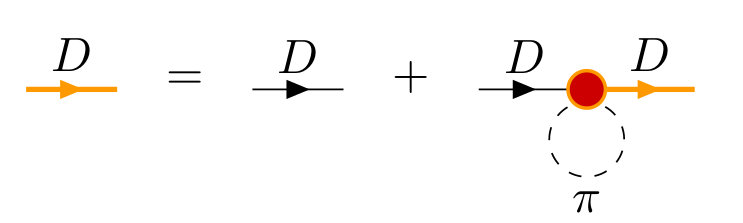}
  \caption{Left: Bethe-Salpeter or $T$-matrix equation at finite temperature for the scattering of $(D + \Phi)_i \rightarrow (D + \Phi)_j$. Right: Dyson equation for $D$-meson propagator at finite temperature. Figures taken from Ref.~\cite{Montana:2020lfi}.}  \label{fig:Tmatrix}
\end{figure}

The scattering amplitude of $D$-mesons off to light mesons is described through a Bethe-Salpeter (or $T$-matrix) equation:
\begin{equation} \red{T_{ij} }  = [ 1- G_{D\Phi} \blue{V}]^{-1}_{ik} \blue{V_{kj}} \ , \label{eq:1}  \end{equation}
where the tree-level perturbative amplitude (also called potentials) $\blue{V_{ij}}$ are temperature-independent functions obtained from the effective Lagrangian at NLO in the chiral expansion. Both potentials and $T$-matrix elements depend on the composition of the incoming and outgoing scattering channels, $i$ and $j$, respectively. Graphically this equation is shown in the left panel of Fig.~\ref{fig:Tmatrix}. The $T$-matrix equation effectively completes the ``unitarization'' procedure, as the new amplitude $T$ exactly satisfies the unitarity constraint.

The two-particle [$(D-\Phi)_k$] loop function $G_{D\Phi}$ accounts for the propagation of both $D$ and $\Phi$ degrees of freedom in the thermal medium. Therefore it explicitly contains spectral information about these two excitations at finite $T$,
\begin{equation} G_{D\Phi}(E,\vec{p};T) =\int\frac{d^3k}{(2\pi)^3}\int d\omega\int d\omega'\frac{ \orange{S_{D}(\omega,\vec{k};T) }S_{\Phi}(\omega',\vec{p}-\vec{k};T)}{E-\omega-\omega'+i\varepsilon} [1+f(\omega,T)+f(\omega',T)] \ . \label{eq:2}  \end{equation}

In this expression---where the external energy has been extended to real values ($+i\varepsilon$ prescription)---the $f$ are the Bose-Einstein distribution functions, and $S(\omega,\vec{k};T)$ are the spectral functions of the different states. Both quantities depend on temperature. The latter can be explicitly written in terms of the $D$-meson self-energy. In view of the known results in the light sector (see discussion within Ref.~\cite{Montana:2020vjg}), we approximate the spectral function of $\Phi$ by the vacuum one, and only consider the spectral modification of the $D$ meson in order to quantify its distortion with respect to the vacuum one. Calling $\green{\Pi_{D}(\omega,\vec{k};T)}$ to the $D$-meson self-energy one has, 
\begin{equation} \orange{S_{D}(\omega,\vec{k};T)}  =-\frac{1}{\pi}{\rm Im\,}\left( \frac{1}{\omega^2-\vec{k}\,^2-m_{D}^2-\green{\Pi_{D}(\omega,\vec{k};T)} } \right) \ ,  \label{eq:3}  \end{equation}
where the self-energy is fixed by the interactions of the $D$-meson with the particles of the medium, mostly pions, and in lesser extent, $K, \bar{K}$ and $\eta$ mesons. Specifically for a light meson $\Phi$ it reads
\begin{equation} \green{\Pi_D(\omega_n,\vec{k};T)}  =T\int\frac{d^3p}{(2\pi)^3}\sum_m\mathcal{D}_\Phi(\omega_m-\omega_n,\vec{p}-\vec{k}\,) \red{T_{D\Phi, D\Phi}(\omega_m,\vec{p}\,)} \ , 
\label{eq:4} \end{equation}
where $\mathcal{D}_\Phi(\omega_m-\omega_n,\vec{p}-\vec{k})$ is the (free) propagator of the light particle. Graphically, the Dyson equation for the light-meson propagator is represented in the right panel of Fig.~\ref{fig:Tmatrix}.

It is important to notice that the set of Eqs.~(\ref{eq:1},\ref{eq:2},\ref{eq:3},\ref{eq:4}) requires a self-consistent solution, as they are coupled among themselves. We attack them in an iterative scheme~\cite{Montana:2020lfi}.
 
 \begin{figure}[ht]
  \centering
  \includegraphics[width=95mm]{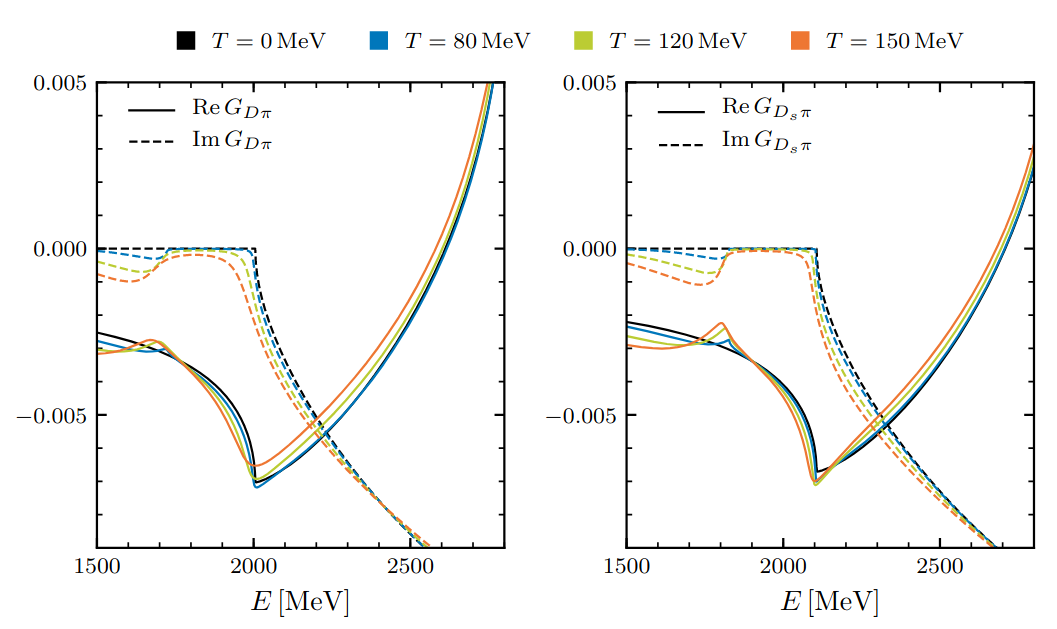}
  \caption{Real and imaginary parts of the two-propagator function with $D$ and $\pi$ states at finite temperature. Left panel: $D\pi$ case. Right panel: $D_s \pi$ case.}
  \label{fig:Gfun}
\end{figure}

  In Fig.~\ref{fig:Gfun} we present the results for the two-propagator function for $D\pi$ (left panel) and $D_s\pi$ (right panel) as functions of the temperature. In particular, the imaginary part of the $G_{D\Phi}$ function plays a key role in the description of the possible open scattering channels, and the generation of a thermal decay width (see later). What is important to recognize is that in vacuum ($T=0$, or black lines in the plots) the imaginary part of this function becomes nonzero above the unitary (or elastic) threshold (for $E=m_D+m_\pi$ and $E=m_{D_s}+m_\pi$, respectively). This characterizes the energy domain where the two particles in the loop can be put on shell. At finite temperatures, a second domain with nonzero values opens below this threshold. We denote that domain the Landau cut, and, when computed with particles on their mass shell, it starts at $E=0$ and goes until the difference of masses (e.g. $|m_D-m_\pi$|). In the self-consistent solution of the equation (with fully off-shell $D$ mesons) the boundaries of the cuts are smoothed, but they are still clearly visible in Fig.~\ref{fig:Gfun}. The contribution of the Landau cut of the $G_{D\Phi}$ function will have a key importance in the calculation of the transport coefficients, as we will see later.
  
\begin{figure}[ht]
  \centering
    \includegraphics[width=100mm]{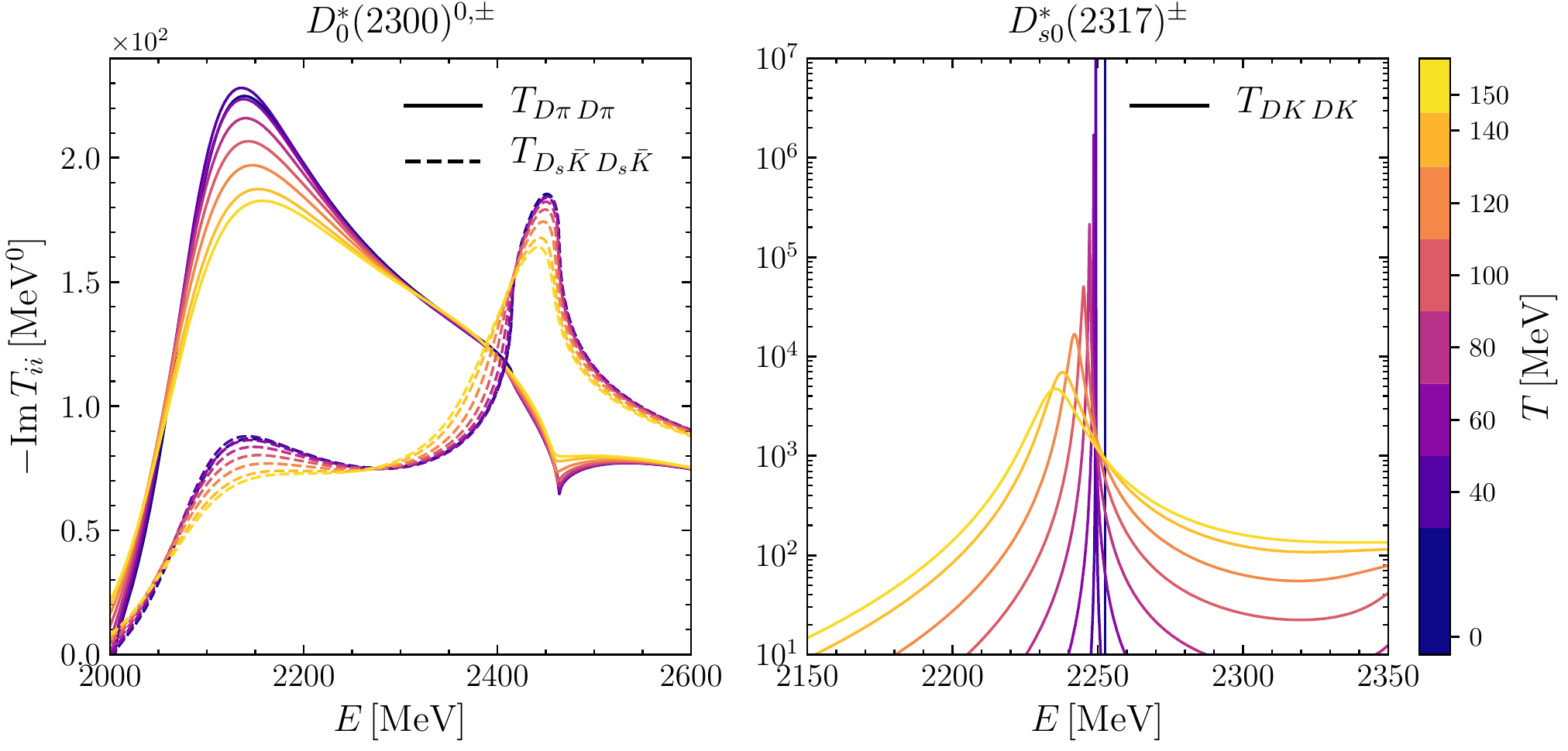}
  \caption{Imaginary parts of the $T$-matrix elements in the diagonal channels. Left panel: $(S,I)=(0,1/2)$ channel. Right panel: $(S,I)=(1,0)$ channel. Figures taken from Ref.~\cite{Montana:2020lfi}.}
  \label{fig:ImT}
\end{figure}

  Figure~\ref{fig:ImT} shows the imaginary part of the resulting $T$-matrix along the real axis in the diagonal channels $D\pi$, $D_s \bar{K}$ (left panel) and $DK$ (right panel). In the first case one can observe the two peaks corresponding to the two-pole structure of the $D_0^* (2300)$ resonance, and its weak dependence with temperature. The lower pole would be clearly visible in the $D\pi$ decay channel, whereas to detect the upper pole one should look into the $D_s \bar{K}$ final state (although its form becomes rather distorted due to the close presence of the elastic thresholds). In the right panel the $D_{s0}^* (2317)$ state is shown as a function of $T$. Notice that this generated state can be characterized as a bound state strictly only at $T=0$, receiving a (very small) thermal width as long as temperature is increased. The numerical values of the masses and decay widths are given in Refs.~\cite{Montana:2020lfi,Montana:2020vjg}.
   
\begin{figure}[ht]
  \centering
  \includegraphics[width=120mm]{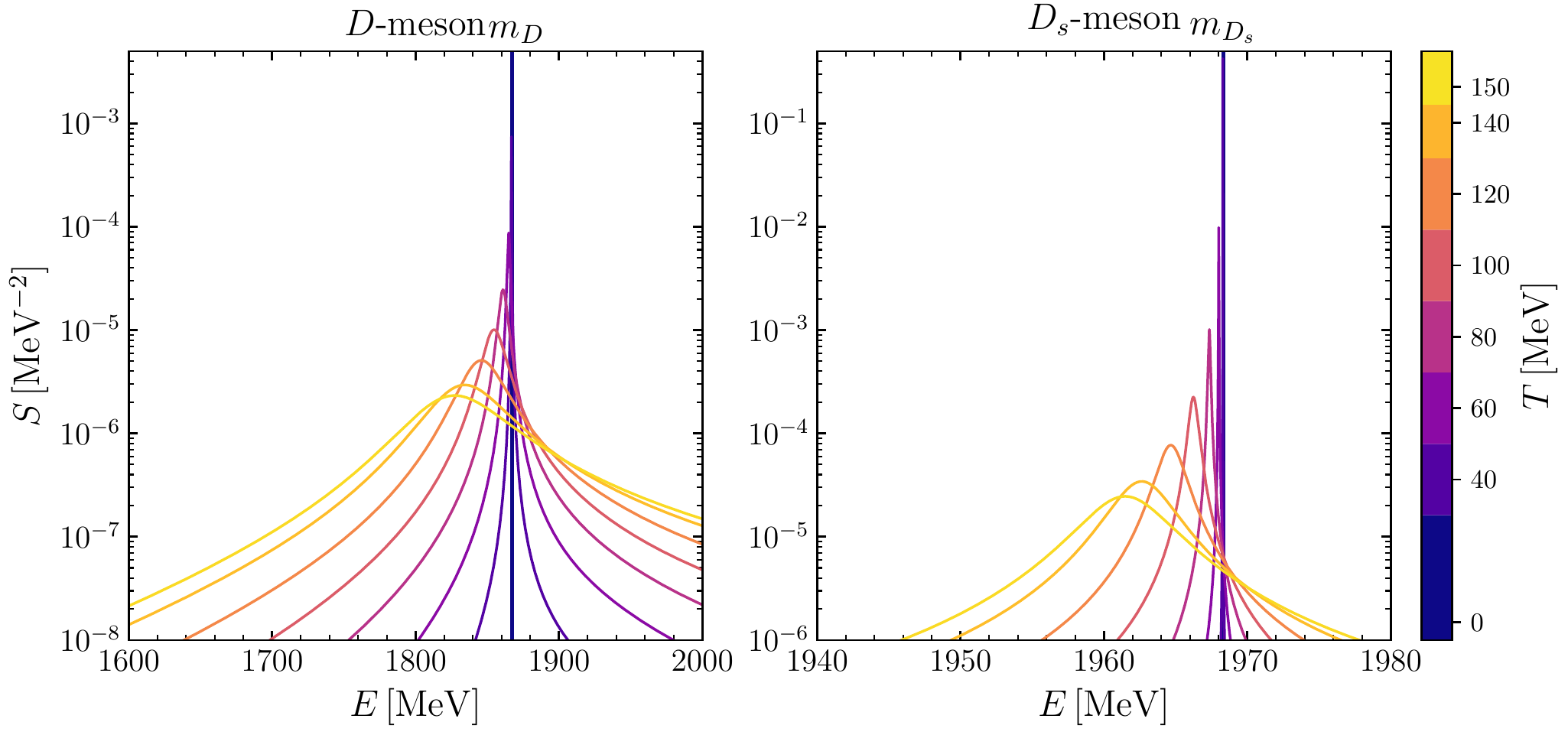}
  \caption{Spectral functions of $D$ (left panel) and $D_s$ (right panel) mesons at rest as functions of the energy and temperature. Figures taken from Ref.~\cite{Montana:2020lfi}.}
  \label{fig:spectral}
\end{figure}

  Also the ground states $D$ and $D_s$ mesons, which are stable degrees of freedom at $T=0$, gain thermal widths at finite temperature. Their spectral functions are shown in the left and right panels of Fig.~\ref{fig:spectral}, respectively.

\begin{figure}[ht]
  \centering
  \includegraphics[width=60mm]{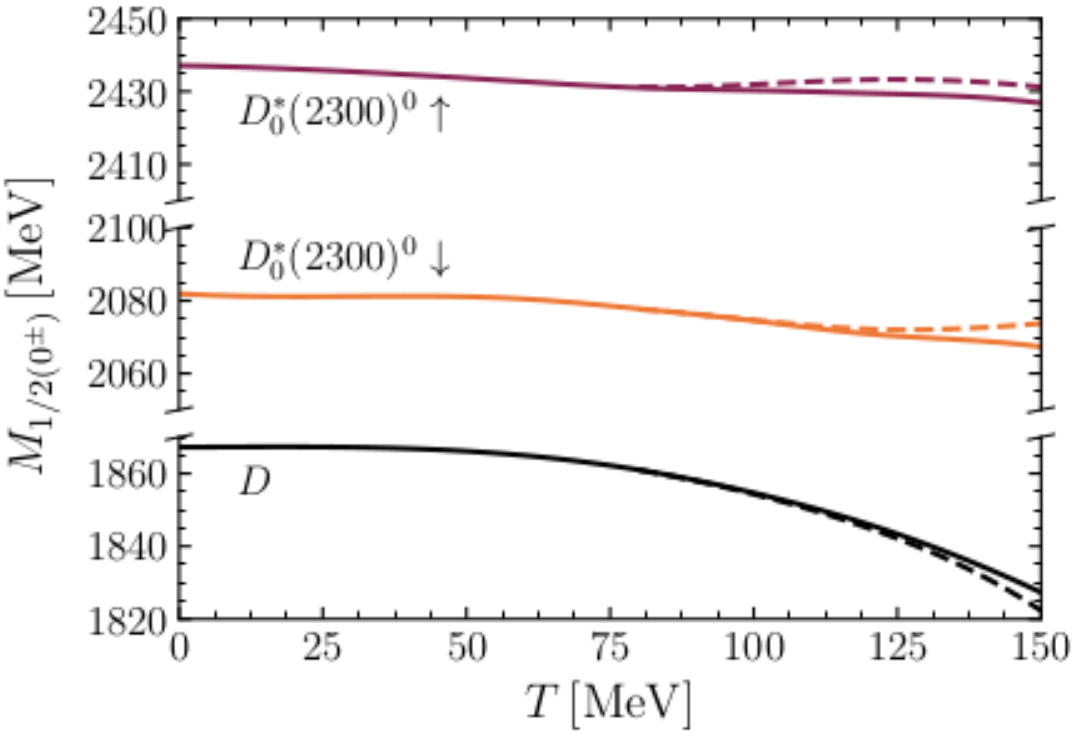}
  \hspace{10mm}
    \includegraphics[width=60mm]{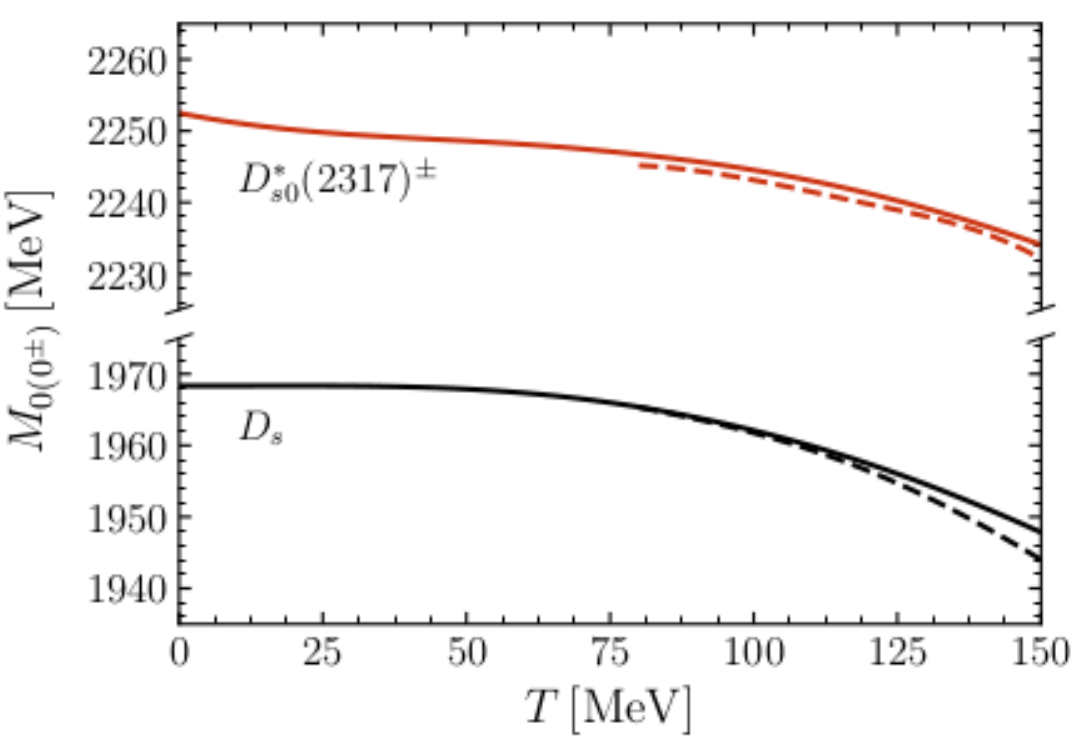}
  \caption{$D$-meson thermal mass as a function of temperature. Left panel: $(S,I)=(0,1/2)$ channel with $D$ ground state ($J^P=0^-$) and two poles of the $D_0^*(2300)$ resonance ($J^P=0^+$). Right panel: $(S,I)=(1,0)$ channels with $D_s$ ground state ($J^P=0^-$) and $D_{s0}^* (2317)$ bound states ($J^P=0^+$). Figures taken from Ref.~\cite{Montana:2020vjg}.}
  \label{fig:mass}
\end{figure}

  The width of these spectral functions is a measure of the thermal contribution to the decay of the $D/D_s$ mesons due to their interactions with particles of the medium. The thermal width monotonically increases with temperature. On the other hand, in the quasiparticle approximation one can obtain the quasiparticle masses (thermal masses modified by the medium) from the location of the peaks. The thermal masses as functions of the temperature can be seen in Fig.~\ref{fig:mass}, where we also include the masses of the two poles of the $D_{s0}^* (2300)$ resonance, and the $D_s^*(2317)$ state.  
  
  All masses shown in Fig.~\ref{fig:mass} present a timid decrease with temperature, being the ground states (black lines) the most affected ones. The effect of $K$ and $\bar{K}$ in the self-energy is shown in dashed lines, and it amounts to a very small correction even at $T=150$ MeV. 
  
  Taking into account that the states considered can be recognized as chiral partners of opposed parity [$D \leftrightarrow D_0^*(2300)$ and $D_s \leftrightarrow D_{s0}^* (2317)$], one may wonder about a possible mass degeneracy at high temperatures, once the chiral restoration temperature around $T_c \simeq 156$ MeV is approached. However, we cannot offer a conclusive answer on this question because in our model---lacking from any information about a phase transition---the temperature dependence is very weak. In addition, no back-reaction of the quark condensate (order parameter of the chiral transition) on the $D$-mesons is implemented. If the expected decrease of the quark condensate is considered to model the temperature dependence of the pion decay constant $f_\pi (T)$ (e.g. via the thermal version of the Gell-Mann-Oakes-Renner relation), then indications of a plausible degeneration of masses can be obtained~\cite{Torres-Rincon:2021wyd}.
  
  In Refs.~\cite{Montana:2020lfi,Montana:2020vjg} we also provide full details on the thermal widths of the different states and their dependence on the light mesons used to dress the heavy-meson propagators. Such a thermal width becomes important in quantifying the absorption/generation rates of these states in the thermal medium. They are also responsible of providing a quasiparticle picture for the charm states, which needs to be incorporated in the kinetic description. This particular aspect, together with the determination of the transport coefficients, is described in the next section.

\section{Off-shell kinetic theory for $D$ mesons} 
  
  The second part of our contribution addresses the non equilibrium description of these heavy systems. Such an approach is needed to define and to address the computation of transport coefficients. Eventually, the calculation of the dissipative coefficients requires only equilibrium properties, as those presented in the previous section. The details concerning the transport theory can be found in our recent work~\cite{Torres-Rincon:2021yga}.      
  
  We start by considering a kinetic theory for the charm distribution function. For this aim, we place the quantum fields of the degrees of freedom along the so-called Kadanoff-Baym contour and use real-time techniques. Details can be found in~\cite{Torres-Rincon:2021yga}, which was based on the classical Refs.~\cite{kadanoff1962quantum,Botermans:1990qi,Davis:1991zc,Blaizot:1999xk}. The resulting Kadanoff-Baym equations describe the time evolution of the $D$-meson Green functions, keeping the light degrees of freedom in equilibrium thanks to their small mass. After performing a gradient expansion of the Kadanoff-Baym equations and a subsequent Wigner transform, one obtains the following kinetic equation for the Wightman function $G_D^< (X,k)$,
  \begin{align} 
& \left(  k^\mu - \frac{1}{2} \frac{\pa \textrm{Re } \Pi^R (X,k)}{\pa k_\mu} \right) \frac{\pa iG_D^< (X,k)}{\pa X^\mu} =\frac{1}{2}  i\Pi^< (X,k)  iG_D^> (X,k) -\frac{1}{2} i\Pi^> (X,k) iG_D^< (X,k) \ .
\label{eq:almostoffGless}
\end{align}  

\begin{figure}[ht]
  \centering
  \includegraphics[width=85mm]{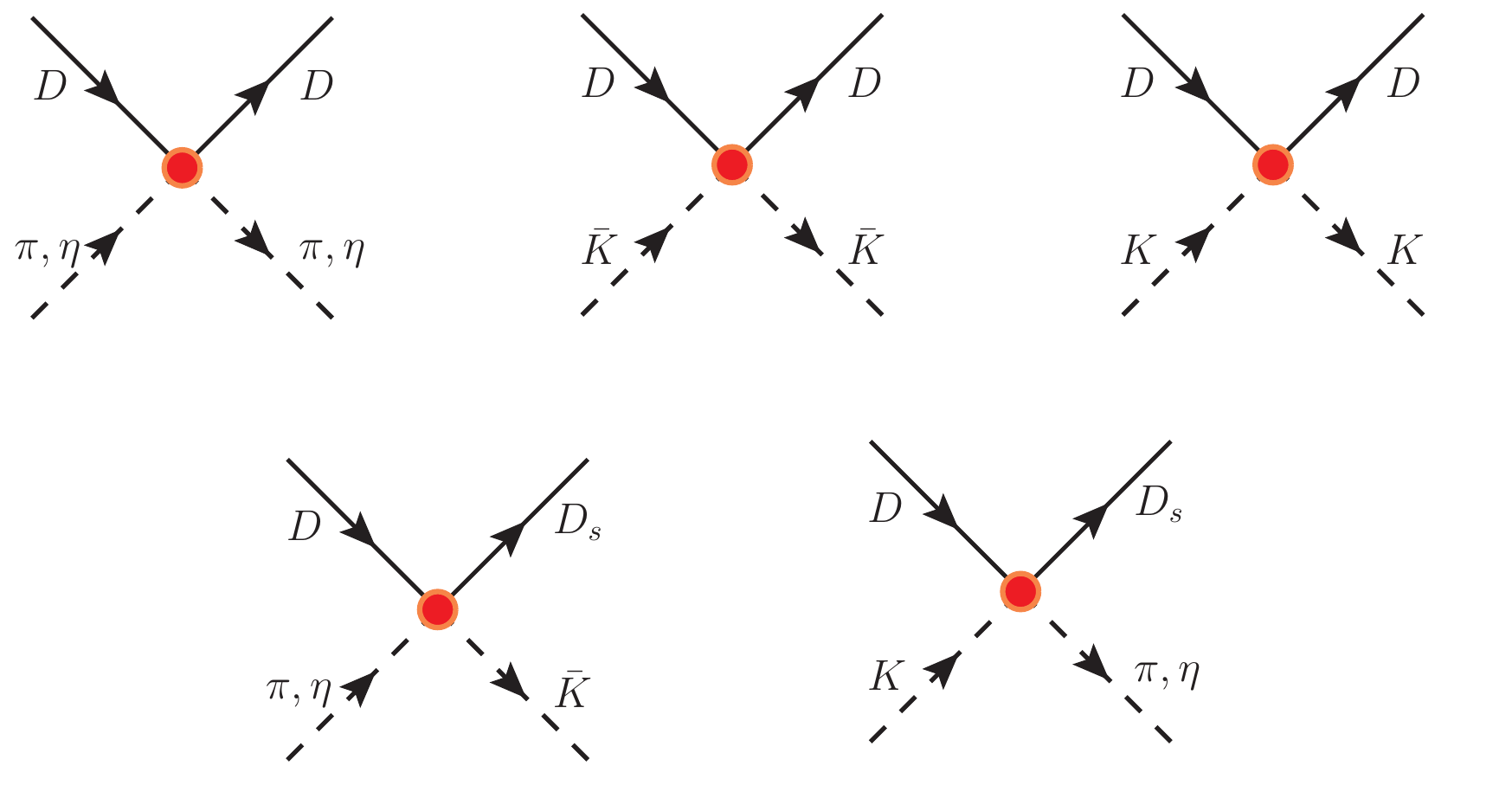}
  \caption{Scattering processes in the collision integral of the Kadanoff-Baym equations under the $T$-matrix approximation. Red circles represent fully retarded $T$-matrix elements.}
  \label{fig:collisions}
\end{figure}

  Several $D$-meson self-energies appear in Eq.~(\ref{eq:almostoffGless}). The $\Pi^R(X,k)$ is the non equilibrium, retarded generalization of Eq.~(\ref{eq:4}). The greater and lesser self-energies $\Pi^< (X,k)$ and $\Pi^> (X,k)$ are related to the gain and loss terms of the collision integral. To close the kinetic equation one should specify the form of these self-energies. For this, we employ---consistently with the approximation used in equilibrium---the so-called $T$-matrix approximation~\cite{kadanoff1962quantum,Danielewicz:1982kk,Botermans:1990qi}. The resulting collision integrals contain a sum over the allowed scattering channels with the interaction vertices being the resumed $T$-matrix elements of Fig.~\ref{fig:Tmatrix}. In particular the possible scattering processes appear in Fig.~\ref{fig:collisions}.
  
   Not all scattering elements are equally important in size, and we have restricted to the elastic processes which provide the main contribution to the transport coefficients. However, it shouldn't be forgotten that to fulfill exact optical theorem (unitarity) all (elastic+inelastic) processes must be accounted for. Another approximation can be made by using the separation of scales between the heavy-meson masses and the other scales in the problem. It allows to expand the collision integrals to eventually arrive to a Fokker-Planck equation~\cite{lifshitz1981physical,Svetitsky:1987gq}. In our case the resulting equation retains the off-shell nature of the $D$ mesons, present already at equilibrium. The final equation can be written as
   
\begin{equation}
\frac{\pa}{\pa t} G_D^< (t,k) = \frac{\pa}{\pa k^i} \left\{ \purple{ \hat{A} (k;T) } k^i G_D^< (t,k) + \frac{\pa}{\pa k^j} \left[ \blue{\hat{B}_0(k;T) } \Delta^{ij} + \orange{ \hat{B}_1(k;T)} \frac{k^i k^j}{{\bf k}^2} \right] G_D^< (t,k) \right\} \ , \label{eq:FP}
\end{equation}
where $\Delta^{ij}=\delta^{ij}-k^ik^j/{\bf k}^2$ and $i,j$ are spatial indices. This equation controls the time evolution of the Wightman function $G_D^<$ according to the first two moments of the collision operator i.e. the drag and the diffusion contributions. In the so-called ``Kadanoff-Baym Ansatz'' $G_D^<$ is the product of the (off-equilibrium) spectral and distribution functions of the $D$ meson. In Eq.~(\ref{eq:FP}) the drag force, transverse and longitudinal diffusion coefficients have been defined~\cite{Torres-Rincon:2021yga},
  
 \begin{equation} \left\{ 
\begin{array}{ccc}
\purple{ \hat{A} (k^0, {\bf k};T) } & \equiv &\left\langle1 - \frac{{\bf k} \cdot {\bf k}_1}{{\bf k}^2} \right\rangle  \\
\blue{  \hat{B}_0 (k^0, {\bf k};T) }  & \equiv&  \frac14 \left\langle {\bf k}_1^2 - \frac{({\bf k} \cdot {\bf k}_1)^2}{{\bf k}^2} \right\rangle  \\ 
\orange{  \hat{B}_1 (k^0, {\bf k};T) }& \equiv& \frac12 \left\langle \frac{[ {\bf k} \cdot ({\bf k}-{\bf k}_1) )]^2}{{\bf k}^2} \right\rangle 
\end{array} \right. \ , 
\end{equation}
where
\begin{align}
\left\langle {\cal F} ( {\bf k}, {\bf k}_1) \right\rangle & \equiv \frac{1}{2k^0} \sum_{\lambda,\lambda'=\pm} \lambda \lambda' \int dk_1^0 \int \prod_{i=1}^3 \frac{d^3k_i}{(2\pi)^3} \frac{1}{2E_22E_3} \orange{S_D(k_1^0,{\bf k}_1)}   (2\pi)^3 \delta^{(3)} ({\bf k}+{\bf k}_3-{\bf k}_1-{\bf k}_2) \nn \\
& \times (2\pi) \delta (k^0+\lambda' E_3- \lambda E_2-k^0_1)  \red{ |T(k^0+ \lambda' E_3,{\bf k}+{\bf k}_3)|^2}  f^{(0)} (\lambda'E_3) \tilde{f}^{(0)} (\lambda E_2) \ {\cal F} ( {\bf k}, {\bf k}_1) \label{eq:average} \ . 
\end{align}
These coefficients can be computed as long as the thermal spectral function $\orange{S_D(k_1^0,{\bf k}_1)}$ and $T$-matrices $\red{ |T(k^0+ \lambda' E_3,{\bf k}+{\bf k}_3)|^2}$ are known. The notation reflects the collision process $k + k_3 \rightarrow k_1 + k_2$, where $k$ and $k_1$ corresponds to the heavy meson, while $k_2$ and $k_3$ to the light particle.

In Ref.~\cite{Torres-Rincon:2021yga} we provide the transport coefficients as functions of temperature where the heavy meson is taken in the so-called static limit ${\bf k} \simeq 0$, and its energy fixed at the quasiparticle mass $k^0=E_{k=0}=m_D(T)$. While the off-shell nature of the $D$-meson precludes a one-to-one correspondence between energy and momentum, the good quasiparticle picture provides a narrow spectral function peaked at $k^0=E_k=\sqrt{k^2+m_D(T)^2}$~(see Fig.~\ref{fig:spectral}). 

In this communication we focus on the static diffusion coefficients, $\kappa (T)$ and $D_s (T)$. The first one is defined in momentum space as
\begin{equation} {\bf \kappa} (T)=2 \blue{B_0 (k^0=E_k, {\bf k}\rightarrow 0;T)} = 2 \orange{B_1 (k^0=E_k,{\bf k}\rightarrow 0;T)} \ , \label{eq:kappa}
\end{equation}
while the diffusion coefficient in coordinate space (multiplied by $2\pi T$) reads
\begin{equation} 2\pi T D_s(T) = \frac{2\pi T^3}{\blue{B_0 (k^0=E_k,{\bf k}\rightarrow 0;T)}} \ . \label{eq:Ds}
\end{equation}

\begin{figure}[ht]
  \centering
  \includegraphics[width=70mm]{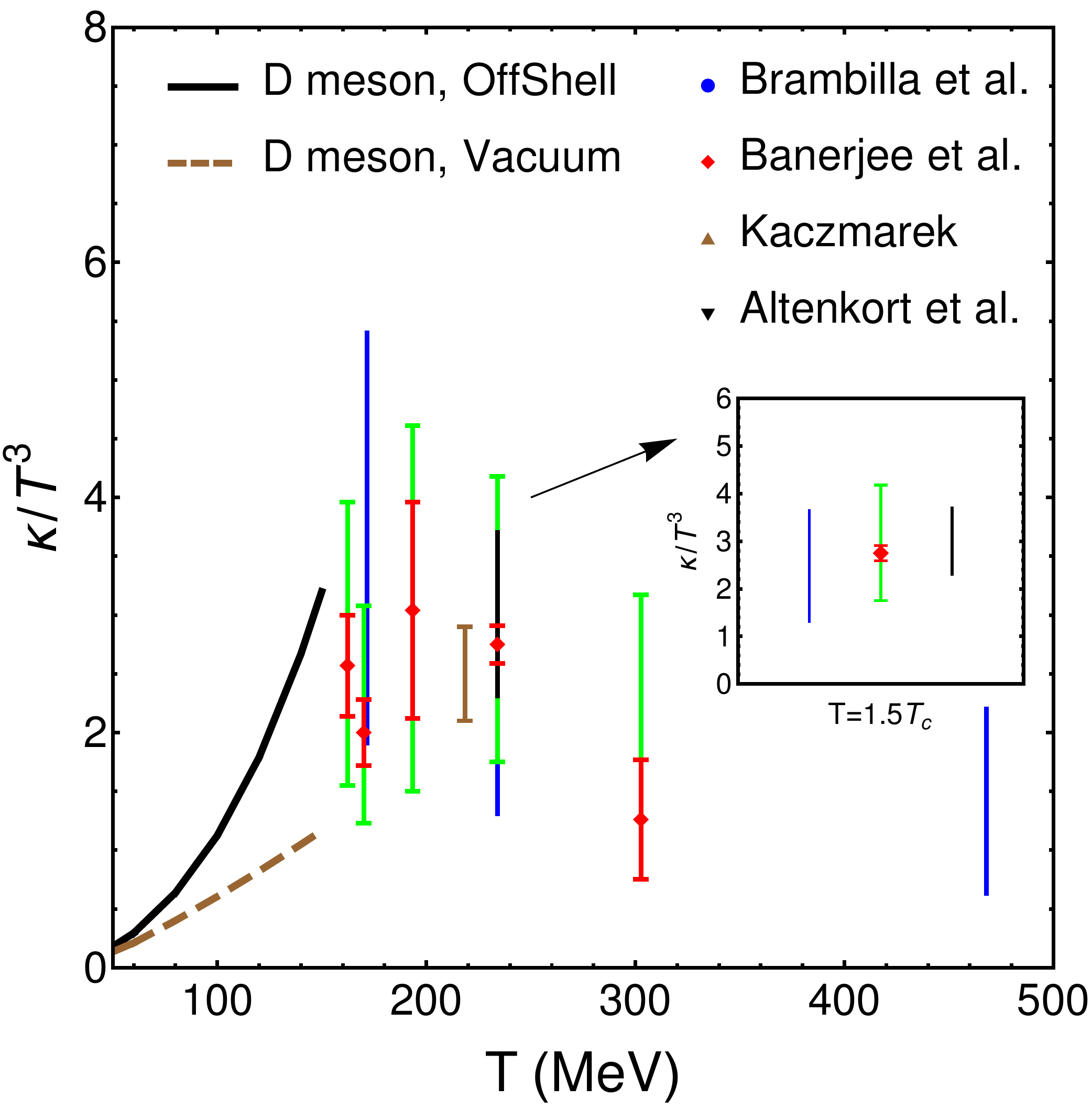}
  \includegraphics[width=70mm]{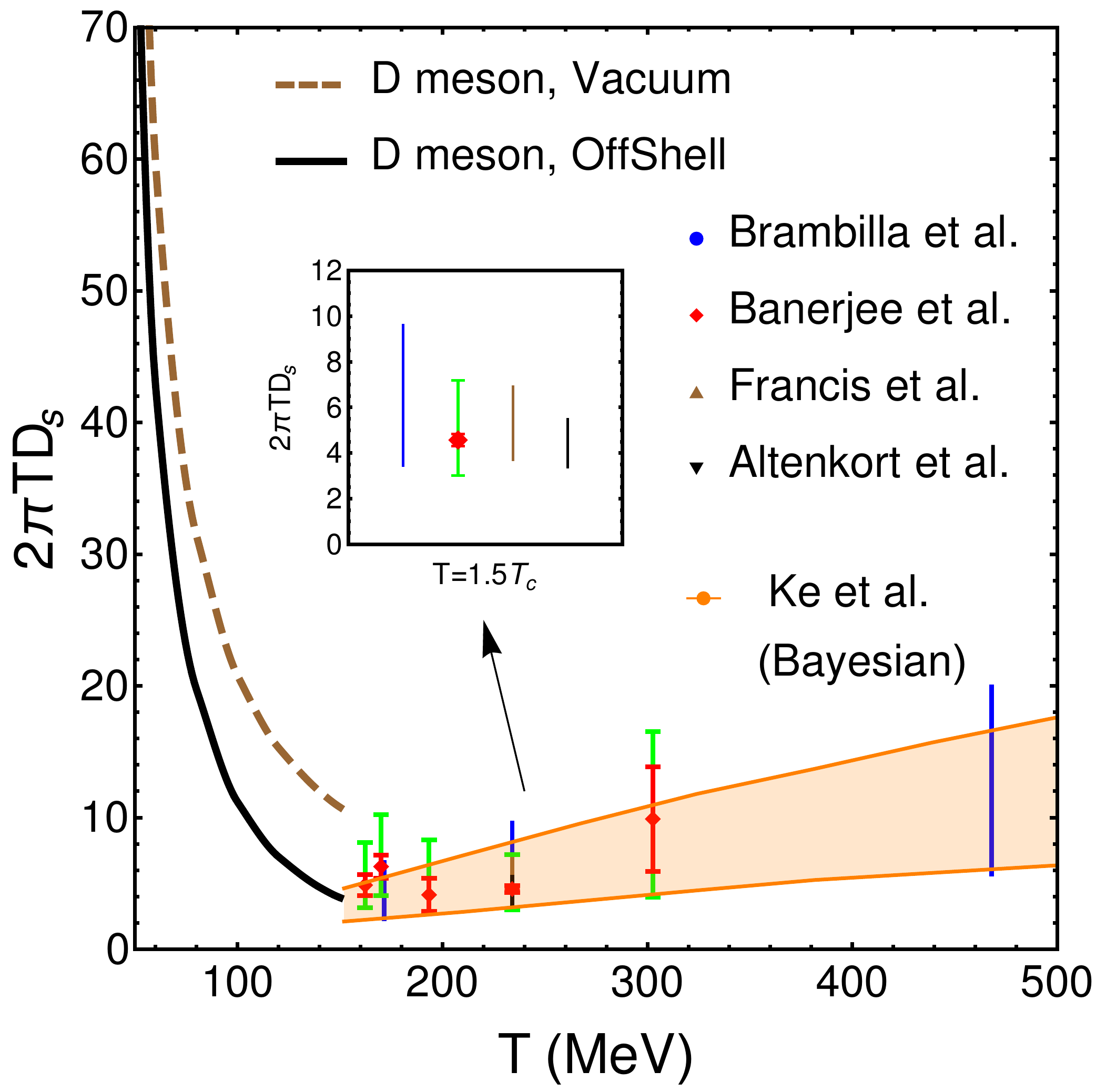}
  \caption{Charm diffusion coefficients in momentum (left panel) and coordinate space (right panel) as functions of temperature. Hadronic results from this work are shown at low temperature while different lattice-QCD calculations, and the posterior results from Bayesian analyses of heavy-ion collisions are shown at high temperatures. Details are provided in Ref.~\cite{Torres-Rincon:2021yga}. Figures taken from this reference.}
  \label{fig:results}
\end{figure}

Both coefficients have been computed in recent lattice-QCD calculations above $T_c$. The latter coefficient has also been extracted from Bayesian analyses of relativistic heavy-ion collisions at high energies. We refer the reader to~\cite{Torres-Rincon:2021yga} for more details and pertinent references. The two diffusion coefficients are shown in Fig.~\ref{fig:results}.

  The dashed brown lines at low temperatures are our calculations when the $T$-matrices employed are those calculated at $T=0$. In addition, the quasiparticle masses are those in vacuum, and no thermal widths are incorporated. That calculation would correspond to the approximations used in previous works~\cite{Tolos:2013kva} (with small differences in the effective interaction due to changes in the low-energy constants and the increase of the coupled-channel basis). The black solid line is our full calculation where thermal effects are taken in a self-consistent way. The considerable increase (decrease) of the $\kappa$ ($2\pi TD_s$) coefficient comes from the new channels appearing when off-shell effects are considered: 
  In Eq.~(\ref{eq:average}) a term with $\lambda'=-1$ evaluates the $T$-matrices below the elastic threshold. From the exact unitarity condition, to evaluate a scattering process below this threshold requires the imaginary part of the 2-particle propagator $G_{D\Phi}$ at the same energy. While at $T=0$ such a contribution would be zero (see how the dashed black line of Fig.~\ref{fig:Gfun}---as opposed to the colored ones---exactly vanishes below the elastic threshold), at finite $T$ the imaginary part develops the Landau cut providing a new contribution to the decay width, and therefore, to the $\lambda'=-1$ term of Eq.~(\ref{eq:average}). According to Eqs.~(\ref{eq:kappa}) and (\ref{eq:Ds}), this positive contribution to the $\purple{A}, \blue{B_0}$ and $\orange{B_1}$ coefficients, translates into the observed increase of $\kappa$ and the decrease of $D_s$. Of course, this contribution from the Landau cut of $G_{D\Phi}$ disappears when $T \rightarrow 0$, as expected.   

It is appealling to see how the corrections from thermal and off-shell effects bring the vacuum results closer to the results above $T_c$ from lattice-QCD calculations and Bayesian studies of heavy-ion collisions. Being the chiral transition known to be a crossover, one expects that the transport coefficients also present a smooth transition at $T_c$, as seen in our results of Fig.~\ref{fig:results}.      
  
\section{Summary}

  We have presented the latest results on $D$ mesons at finite temperature including effective interactions, generation of dynamical states, thermal masses and widths, kinetic equation and transport coefficients. We have quantified the degree of modification that the temperature causes on these states. In particular, we have stressed that a thermal width appears for all heavy degrees of freedom, making the stable states in vacuum to gain certain decay width when they are immersed in a bath. Despite these thermal widths, the quasiparticle picture remains an excellent approximation for the temperatures considered in this work, and the medium modifications of the masses are always relatively small. Using the Kadanoff-Baym approach, a derivation of an off-shell kinetic equation and its reduction to an off-shell Fokker-Planck has been described. Finally, we have addressed the transport coefficients of these systems, and found that the off-shell extension brings important modifications of the dissipative coefficients when both thermal and off-shell effects are introduced. A better matching around $T_c$ with several results at high temperatures has been obtained, providing indications of improvement after incorporating these new effects.         

\section{Acknowledgments}
    
J.M.T.-R. and L.T acknowledge support from the Deutsche Forschungsgemeinschaft through projects no. 411563442 and no. 315477589 - TRR 211.
G.M. acknowledges support from the FPU17/04910 Doctoral Grant from the Ministerio de Educaci\'on, Cultura y Deporte. G.M. and A.R. acknowledge support from the Ministerio de Economía y Competitividad under the project MDM-2014-0369 of ICCUB and from the Ministerio de Ciencia e Innovaci\'on under the project PID2020-118758GB-I00. In addition, the research of L.T. has been supported by the Ministerio de Ciencia e Innovaci\'on and the European Regional Development Fund (ERDF) under contract PID2019-110165GB-I00 (Ref.10.13039/501100011033), by Generalitat Valenciana under contract PROMETEO/2020/023  and by the EU STRONG-2020 project under the program  H2020-INFRAIA-2018-1, grant agreement no. 824093.

\end{document}